# Manuscript Template

## Exploration on the Two-stream Instability in the Polar Cusp Under Solar Storm Disturbances and its Potential Impacts on Spacecraft

Ji-Kai Sun[1], Lei Chang[1*], Yu Liu[2], Guo-Jun Wang[3], Zi-Chen Kan[1], Shi-Jie Zhang[1], Jing-Jing Ma[1], Ding-Zhou Li[1], Ying-Xin Zhao[1]

[1] School of Electrical Engineering, Chongqing University, Chongqing 400044, China.
[2] CAS Key Laboratory of Geospace Environment, School of Earth and Space Sciences, University of Science and Technology of China, Hefei 230026, China.
[3] State Key Laboratory of Space Weather, National Space Science Center, Chinese Academy of Sciences, Beijing 100190, China.

*E-mail: leichang@cqu.edu.cn

**Abstract.** During solar storms, the polar cusp often exhibits electron populations with distinct velocity distributions, which may be associated with the two-stream instability. This study reveals the evolution of the two-stream instability associated with electron velocities and the interaction between the growth phase of the two-stream instability and the electrostatic solitary waves (ESWs). The results from particle-in-cell (PIC) simulations are compared with satellite observational data and computational outcomes. The potential risks associated with two-stream instability, including surface charge accumulation and communication system interference on spacecraft, are also explored. The findings show that, in the high-latitude polar cusp region, the interaction between the solar wind plasma propagating along magnetic field lines and the upward-moving ionospheric plasma could drive two-stream instability, leading to the formation of electron hole structures in phase space and triggering a bipolar distribution of ESWs. When the spatial magnetic field and wave vector meet specific conditions, the enhanced electron cyclotron motion could suppress the formation of two-stream instability and electron hole structures, leading to a reduction in the amplitude of the ESWs. The results offer valuable insights for a deeper understanding of the impact of solar storms on the polar cusp environment, as well as for monitoring electromagnetic environment and ensuring the stable operation of spacecraft.

## 1. INTRODUCTION

Solar storms are intense eruptive events on the Sun that cause disturbances within the heliosphere. These events primarily release enhanced electromagnetic radiation, high-energy charged particles, and plasma clouds. The solar wind, which serves as the medium for the propagation of solar storms, is primarily composed of protons, electrons, α-particles, neutral particles, and heavy ions, with protons and electrons constituting the dominant components of the plasma [1-2]. Solar wind plasma is typically considered as collisionless plasma, where collective behaviors arise through wave-particle interactions. These interactions excite phenomena such as electrostatic solitary waves (ESWs) [3], whistler waves [4], and ion acoustic waves [5], which reflect the dynamic nature of solar wind plasma. The polar cusp, serving as the natural gateway between Earth and space, plays a crucial role in solar wind research. The schematic illustration of the solar wind entering the polar cusp is shown in Figure 1. Solar wind mass, momentum, and energy can directly enter Earth's magnetosphere through the polar cusp, without crossing magnetic field lines [6-8]. Therefore, studying the solar wind in this region provides a more accurate understanding of space weather variations and offers valuable early warning information for the stable operation of spacecraft in orbit.

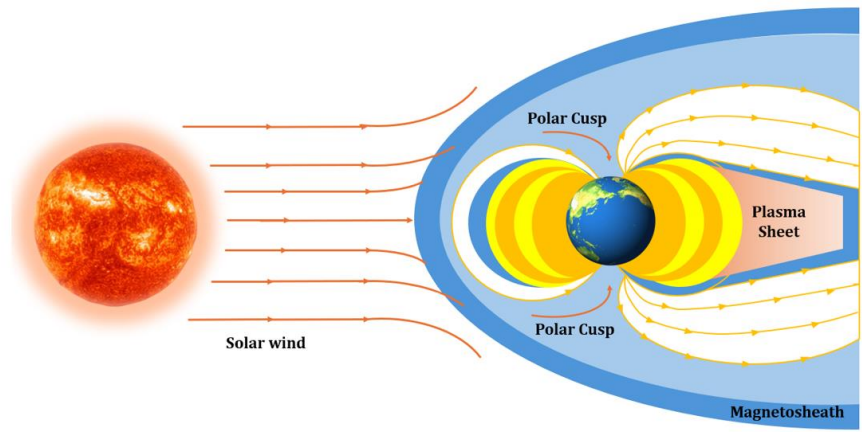

**Fig. 1.** The interaction of solar wind with earth's magnetosphere and plasma layer

In the polar cusp plasma environment, the coexistence of plasma populations with different sources and energy characteristics is a common phenomenon. During solar storms, the ionospheric plasma gains additional kinetic energy. The Earth's magnetic field converges in the polar regions, causing the plasma to concentrate and ascend toward the polar cusp along the magnetic field lines. The solar wind plasma and the upward-moving ionospheric plasma converge in the polar cusp, forming two distinct particle populations with different energy levels. When the drift velocity difference between these populations becomes sufficiently large, energy exchange and wave excitation are enhanced, potentially triggering two-stream instability [9-10]. In addition to the polar cusp, two-stream instability can also occur in regions such as the magnetosheath [11], Earth's bow shock [12], and auroral acceleration zones [13]. Two-stream instability is associated with ESWs and phase-space electron holes [14-16]. Electron holes in electrostatic modes are regarded as static Bernstein-Greene-Kruskal (BGK) solutions to the Vlasov-Poisson equations, reflecting nonlinear interactions between particles and electromagnetic waves in plasma [17]. Currently, the Cluster, Defense Meteorological Satellite Program (DMSP), and Polar satellites have conducted extensive observations of the polar regions, focusing on solar wind plasma at different altitudes. The findings from the Polar satellite confirmed the presence of ESWs in the polar region, along with the localized positive potential structure of electron holes. However, further investigations are required to fully understand the temporal and spatial variations of ESWs [18]. Except for the magnetosheath, ESWs in all other regions are excited by two-stream instability, reflecting the spatiotemporal evolution and decay of plasma beams [19]. The small Debye length, however, poses significant challenges in detecting the pulse structure of ESWs [20-21]. Therefore, we employed simulation methods to study two-stream instability and ESWs excitation. These results contribute to a deeper understanding of the physical mechanisms governing the excitation of two-stream instability.

Research on the triggering mechanisms of two-stream instability and ESWs in the polar cusp during solar storms remains limited. Few studies have addressed the potential negative impacts of space plasma two-stream instability, such as surface charge accumulation on spacecraft and interference with communication signals. Considering the magnetic field distribution and spatial variations in the polar cusp, we develop a model for particle interactions based on the BGK solution of the Vlasov–Poisson equations, employing a combined approach of quantitative calculations and qualitative analyses. The excitation of the two-stream instability and the characteristics of wave propagation are characterized through the dispersion relation, electron density distribution, and velocity distribution. Moreover, we conduct a qualitative assessment of the potential risks associated with spacecraft surface charging and communication system interference. To the best of our knowledge, this study is the first to combine the triggering mechanisms of two-stream instability in the polar cusp with its potential adverse effects.

The structure of this paper is organized as follows: In the "Model and Implementation" section, we primarily present the governing equations and the numerical scheme for two-stream instability. Furthermore, we define the physical parameters of the upward-moving ionosphere electrons and downward-moving solar wind electrons in the polar cusp. In the "Results and Analysis" section, we analyze the evolution of the two-stream instability and wave characteristics, and further study the impact of the magnetic field on the suppression of instability. Finally, the "Conclusion" section summarizes the findings and provides recommendations for future research.

## 2. MODEL AND IMPLEMENTATION

Cluster satellite data suggest that the ratio of magnetic field fluctuations to electric field fluctuations $\delta B/\delta E$ in the polar cusp is approximately $10^{-3}$, indicating that they can be approximated as electrostatic structures. Two-stream instability induces local electron density variations and electric field fluctuations through the interaction between particle streams, which then evolve into ESWs through nonlinear processes. While ESWs exhibit some electromagnetic wave characteristics, the ratio between magnetic field fluctuations and electric field fluctuations is small, indicating that the electrostatic properties dominate [22-23]. The Vlasov-Poisson equations provide a fundamental framework for describing the spatiotemporal evolution of plasmas and the propagation of electromagnetic waves. However, due to the inherent nonlinearity of the equations, their solution typically requires the use of numerical methods, such as particle-in-cell (PIC) simulations. The BGK solution of the Vlasov-Poisson equations provides an effective method for describing ESWs and phase-space electron hole phenomena in plasma [24-25]. The BGK solution maintains a self-consistent relationship between particles and the electric field, ensuring a more accurate model for the interaction between solar wind plasma and upward-moving ionospheric plasma. Building upon this, we further derive the dispersion relation between solar wind electrons with different temperatures and drift velocities and upward-moving ionosphere electrons in the polar cusp, and discuss the composition of the BGK solution. The BGK solution exists only in the electrostatic mode. Therefore, in the modeling and solution process, we primarily use the analytically solvable Vlasov-Poisson model. In the Vlasov-Maxwell model, the magnetic field perturbations in the electromagnetic mode disrupt the energy conservation, preventing the derivation of an analytically solvable BGK solution [26]. Consequently, only numerical approximations, which are computationally expensive, can be used for analysis. To balance model accuracy and simulation analysis, we mainly employ electrostatic mode theory for the modeling process. For simulation analysis, we use PIC simulations to investigate the two-stream instability under both electromagnetic and electrostatic modes, making the simulation results more representative of the actual space conditions.

### 2.1 Theoretical analyses

In the study of two-stream instability in the polar cusp during solar storms, it is typically assumed that collisions between particles and background gases can be neglected. The motion of particles is solely governed by external factors such as electric and magnetic fields. The particle beams are treated as continuous fluids on a macroscopic scale. Due to the complexity of analyzing positive ions, we consider the positive ions as a uniform background, focusing primarily on the motion of electrons. When the relative velocity between the low-energy and high-energy electron beams exceeds a critical threshold, energy transfer occurs between the electron beams. This leads to the radiation of electromagnetic waves and energy transfer, triggering intense two-stream instability.

The essence of the two-stream instability lies in the growth of electrostatic perturbations caused by the interaction between two counter-streaming plasma beams. In this study, we focus on the case where the angle between the beam and the magnetic field is either zero or small. We adopt an electrostatic mode for modeling. The established model will be compared and analyzed with the subsequent simulation results.

Firstly, we defined the two electron populations in the two-stream system. The equilibrium distribution function for the solar wind electrons and the ionospheric up-going electrons is formulated as follows:

$$f_0(v_x) = \frac{n_s}{\sqrt{\pi} v_{ths}} exp\left(-\frac{(v_x - v_s)^2}{v_{ths}^2}\right) + \frac{n_i}{\sqrt{\pi} v_{thi}} exp\left(-\frac{(v_x - v_i)^2}{v_{thi}^2}\right). \qquad (1)$$

the $f_0(v_x)$ represents the initial velocity distribution of two particle populations in the $v_x$ direction. Electron density in the solar wind is denoted by $n_s$, with a horizontal drift velocity along the x-axis $v_s$ and thermal velocity $v_{ths} = \sqrt{k_B T_s/m_e}$. Similarly, for the ionospheric up-going electrons reaching the polar cusp region, the electron density is also $n_i$, with the drift velocity $v_i$ and thermal velocity $v_{thi} = \sqrt{k_B T_i/m_e}$. The electron thermal velocity represents the horizontal thermal velocity of electrons along the x-axis. It is assumed that the positive ions are stationary, and the total electron density is constrained by the relation $n_s+n_i=n_0$, where $n_0$ represents the positive ions.

Secondly, the Vlasov–Poisson equation can intuitively describe the relative drift between two electron beams and the evolution of their perturbations. The Poisson equation describes how the electric field is generated by a charge distribution. The charge distribution induces an electric field in the surrounding space, and the strength and distribution of the electric field and potential are determined by the charge distribution. The Poisson equation is expressed as follows:

$$\frac{\partial E}{\partial x} = \frac{q_e}{\varepsilon_0}\left(n_0 - \int f dv_x\right). \quad (2)$$

The Vlasov equation is a partial differential equation that primarily describes the evolution of particles in phase space over time in the absence of collisions, as shown in the following equation:

$$\frac{\partial f}{\partial t} + v_x \frac{\partial f}{\partial x} - E \cdot \frac{q_e}{m_e} \frac{\partial f}{\partial v_x} = 0. \quad (3)$$

The electron charge $q_e > 0$ represents the absolute value of the charge. The electron mass is represented by $m_e$. The symbol $f$ represents the particle distribution function. The symbol $\varepsilon_0$ represents the vacuum permittivity. In the absence of collisions, the distribution function of particles in phase space is conserved, meaning that the total number of particles remains unchanged as they move from one position to another. In the evolution of the two-stream instability, background electrons gain kinetic energy from the beam electrons, causing the velocity distribution to extend towards higher velocity regions. The perturbations in the particle velocity distribution induced by electron-electron interactions, along with the local electric field, will alter the particle's dynamics: $f = f_0 + f_1 e^{i(kx-\omega t)}$ and $E = E_1 e^{i(kx-\omega t)}$, with $f_0$ represents the initial particle distribution function, while $f_1$ denotes the change in the distribution function induced by disturbances. We linearize the Vlasov equation, which gives the following equation:

$$f_1 = \frac{iq_e E_1}{m_e(\omega - kv_x)} \frac{\partial f_0}{\partial v_x}. \quad (4)$$

Here, the symbol $f_1$ is the explicit solution derived from the first-order Vlasov equation. The resonance denominator $1/(\omega - kv_x)$ indicates velocity-matched particles driving the instability. When the particle velocity is close to the wave's phase velocity, even a tiny electric-field perturbation can be amplified through Landau resonance, driving the plasma unstable.

The purpose of the linear stability analysis for the electron distribution function induced by the perturbation electric is to examine whether the perturbation terms will cause the system to diverge. The perturbation electric and magnetic fields in the equations vary with both time and space. If the perturbation $f_1$ grows exponentially with time (i.e., $f_1 \sim e^{\gamma t}$ and $\gamma>0$), the system is unstable. If the perturbation decays or remains constant (i.e., $\gamma \leq 0$), the system is stable.

Thirdly, in the study of solar storm, the interaction between electrons and electromagnetic waves is a crucial aspect of understanding solar wind dynamics and space weather. Following the work in reference [27], we introduced a plasma response function model based on plasma kinetic theory. This model rigorously incorporates collisionless Landau damping into the fluid framework, thereby achieving the convergence of fluid and collisionless kinetic descriptions. The response characteristics of electrons can be expressed by the following equation:

$$\chi_j = \frac{\omega_{pj}^2}{k^2 v_{thj}^2}[1 + \zeta_j Z(\zeta_j)] \quad (5)$$

$$\zeta_j = \frac{\omega - kv_j}{\sqrt{2}kv_{thj}} \quad (6)$$

$$Z(\zeta_j) = \frac{1}{\sqrt{\pi}} \int_{-\infty}^{\infty} \frac{e^{-t^2}}{t - \zeta_j} dt \ (j = s, i). \quad (7)$$

The polarizability $\chi_j$ describes the response of solar electrons to electromagnetic waves, which characterizes the dynamic interaction between electrons and the wave field. The correction term $\zeta_j Z(\zeta_j)$ in the polarizability accounts for Landau damping and the propagation of electromagnetic waves, reflecting the non-ideal effects in the plasma. The variable $\zeta_j$ normalizes the velocity and wave number units to quantify the difference between the wave frequency and the electron's relative velocity, directly reflecting the energy exchange and dynamical behavior between the electrons and the waves. In practical analysis of the dispersion function, the Fried-Conte function $Z(\zeta_j)$ is expressed as $Z(\zeta_j) = Re[Z(\zeta_j)] + iIm[Z(\zeta_j)]$. Here, the imaginary part represents the Landau resonance term, which can be written as $Im[Z(\zeta_j)] = \sqrt{\pi}e^{-\zeta_j^2}$. The real part, however, must be obtained through principal value integration or series expansion, and its specific form will be provided in the dispersion relation analysis during the simulations. The frequency in the equation is represented as: $\omega_{pj}^2 = e^2 n_j / \epsilon_0 m_e$.

By considering the behavior of both solar wind and ionosphere electrons, we can derive a mathematical model for the dispersion relation to describe their responses. By combining the response characteristics of solar wind and ionosphere electrons, a unified equation can be established to reveal the interaction between the electron populations and electromagnetic waves in the solar wind as follows:

$$\epsilon(k, \omega) = 1 + \chi_s + \chi_i = 0 \tag{8}$$

$$1 + \frac{\omega_{ps}^2}{k^2 v_{ths}^2}[1 + \zeta_s Z(\zeta_s)] + \frac{\omega_{pi}^2}{k^2 v_{thi}^2}[1 + \zeta_i Z(\zeta_i)] = 0. \tag{9}$$

Equation (9) describes the interaction between the propagation characteristics of electromagnetic waves and the electron population, and can be considered as the dispersion relation for the electromagnetic wave. The frequency $\omega$ represents the temporal rate of change of the wave, while the wave number $k$ characterizes the spatial rate of change. The wavelength refers to the distance over which the wave propagates. The dispersion relation reveals how the wave propagates at different wave numbers and frequencies, and provides insights into the plasma's response to external disturbances under varying electromagnetic conditions. The dielectric function $\epsilon(k, \omega)$ is used to describe the response of the medium to the electromagnetic waves. When the frequency $\omega$ of the electromagnetic wave approaches the characteristic frequency of the electron group, the dielectric function $\epsilon(k, \omega)$ will approach zero, indicating that the system enters a resonant state. When $\epsilon(k, \omega)=0$, it indicates that the electromagnetic wave cannot propagate at this frequency and wave number, which corresponds to the root of the dispersion relation.

Subsequently, as the two-stream instability approaches saturation, the system self-organizes into a BGK electron hole. The initially counter-drifting beams collapse into a coherent phase-space vortex; particles are trapped within its separatrix. The resulting equilibrium distribution is a function of the single-particle energy alone:

$$f(x, v_x) = F(\varepsilon) \tag{10}$$

$$\varepsilon = \frac{1}{2}m_e v_x^2 - q_e \varphi(x). \tag{11}$$

The premise of Equation (10) and (11) is that we assume the electron distribution function in steady state depends solely on energy. The single-particle energy is composed of the electron's kinetic energy $(m_e v_x^2)/2$ and electrostatic potential energy $q_e \varphi(x)$. We further construct a detailed description of the bi-thermal distribution for the two beam electron populations based on the assumptions:

$$F(\varepsilon) = \sum_{j=s,i} \frac{n_j}{\sqrt{\pi} v_{thj}} exp\left(-\frac{\varepsilon - \frac{1}{2}m_e v_j^2 + q_e \varphi_0}{k_B T_j}\right) \tag{12}$$

Here, the symbol $\varphi_0$ is the reference electrostatic potential at the center of the potential well. The exponential form of the distribution function in Equation (12) originates from the combination of the Maxwell–Boltzmann distribution and single-particle energy conservation. As the particle energy increases, the probability decreases exponentially with the thermal energy scale $k_B T_j$, while the inclusion of drift kinetic energy and the reference potential ensures the physical consistency of different electron populations in phase-space distributions.

Finally, the Poisson equation reconstruction is central to the self-consistency of BGK electron holes, fundamentally establishing a closed-loop feedback between the electrostatic potential distribution and the trapping in electron phase space. The expression is given as follows:

$$\frac{d^2\varphi}{dx^2} = \frac{q_e}{\epsilon_0}\left(n_0 - \int F(\varepsilon)dv_x\right) \tag{13}$$

The electrostatic potential $\varphi(x)$ determines the phase-space structure of the distribution function $F(\varepsilon)$, and the integral of $F(\varepsilon)$ in turn influences the second derivative of $\varphi$, forming a self-organizing field.

The two-stream instability model, which integrates the Vlasov–Poisson framework, the evolution of BGK electron holes, and the dielectric response function, provides a comprehensive basis for describing wave propagation and wave–particle interactions in plasmas. The development of this instability is governed not only by electron temperature and drift velocity, but also by the detailed phase-space distribution of electrons and their coupling with plasma waves. This model will serve as the theoretical foundation for interpreting the subsequent simulation results.

## 2.2 Numerical computation

The Kyoto University Electromagnetic Particle Code: 1D version (KEMPO1) is primarily used to simulate electromagnetic processes in space plasmas [28]. We utilize KEMPO1 to compute two-stream instability in the present work, which employs the Finite Difference Time Domain (FDTD) method to solve Maxwell's equations, describing the interaction between the electric field E, magnetic field B, and charged particles. The code also effectively simulates the thermal fluctuation effects in the electric field, which is crucial for understanding the instability mechanisms in plasmas. The Buneman-Boris method is one of the commonly used numerical solutions in PIC simulations, where the key concept involves using half-step and full-step updates to enhance the numerical stability of particle motion in electromagnetic fields. In the code, Fourier transforms are applied to analyze the spectral characteristics of the electromagnetic fields, allowing for the extraction of frequency domain information from both time and space domain signals. This enables the analysis of the behavior and propagation characteristics of different wave modes.

In the subsequent simulations, we combine both electrostatic and electromagnetic modes. The electrostatic mode neglects the induced electric field and primarily solves the evolution of the two-stream instability by solving the Poisson equation. In contrast, the electromagnetic mode utilizes the full set of Maxwell's equations, and the subsequent analysis investigates the impact of different velocity directions and electron cyclotron frequency on the two-stream instability. The PIC method is employed to solve the particle motion in the polar cusp two-stream instability under solar storm disturbances. In this approach, particles are treated as discrete entities, and their motion within the spatial electromagnetic field is tracked throughout the simulation. To study wave-particle interactions and the propagation of electromagnetic waves, the FDTD method is used to solve the polar cusp system under solar storm disturbances.

The KEMPO1 code normalizes different parameters, transforming the original system equations into dimensionless form, allowing the simulation scale to be adjusted according to different spatial conditions, thereby achieving stable and efficient computational results. Time and space normalization is a key aspect of the code parameter settings. Time normalization is typically performed using the plasma characteristic time, which represents the time required for a particle to undergo one complete oscillation in the plasma: $t_{norm} = \omega_p t$. Space normalization is usually done through the Debye length: $x_{norm} = x/\lambda_D$, which describes the scale of charge screening in the plasma. If the grid spacing is $x > 2\lambda_D$, non-physical behavior may occur, leading to numerical instability.

During the solution process, two sets of spatial grid systems are defined along the x-axis. The quantities $E_y$、$B_y$、$J_y$ and $\rho_i$ are defined on full integer grids, while $E_x$、$E_z$、$B_z$ and $J_x$ are defined on half-integer grids. Central differences $\Delta x$ and $\Delta t$ are used to approximate the derivative terms in Maxwell's equations. The core idea is to discretize both the spacetime coordinates and the components of the electromagnetic fields, transforming the partial differential equations into difference equations, thus enabling the solution of the electromagnetic field's evolution in the time domain and improving computational efficiency. The discretization of the electric and magnetic field intensities is represented by the following formulas:

$$\frac{E_{x,i+\frac{1}{2}} - E_{x,i-\frac{1}{2}}}{\Delta x} = \frac{\rho_i}{\varepsilon_0} \tag{14}$$

$$\frac{E_{y,i+1} - E_{y,i}}{\Delta x} = \frac{\partial B_{z,i+\frac{1}{2}}}{\partial t} \tag{15}$$

$$\frac{E_{z,i+1/2} - E_{z,i-1/2}}{\Delta x} = \frac{\partial B_{y,i}}{\partial t} \tag{16}$$

$$\frac{B_{y,i+1} - B_{y,i}}{\Delta x} = \mu_0 J_{z,i+\frac{1}{2}} \tag{17}$$

$$\frac{B_{z,i+1/2} - B_{z,i-1/2}}{\Delta x} = -\mu_0 J_{y,i}. \tag{18}$$

Here, the symbol $\mu_0$ represents the magnetic permeability of free space. When solving for the set parameters, the time step and spatial step must satisfy the Courant condition $c\Delta t < \Delta x$ to ensure the convergence and stability of the numerical solution. We map the particle motion information onto the grid, and based on the particle positions and charge distribution, the charge density at each grid point can be calculated. Additionally, the current density can be computed using the particles' velocity and position.

$$\rho_i = \frac{1}{\Delta x} \sum_j q_j W(x_j - X_i) \tag{19}$$

$$J_i = \sum_j q_j v_{x,j} W(x_j - X_i). \tag{20}$$

Here, the symbol $\rho_i$ represents the charge density at grid point $i$, which is the total charge per unit volume. $q_j$ represents the charge of particle $j$, $J_i$ represents the current density at grid point $i$, $v_{x,j}$ is the velocity component of particle $j$ in the $x$-axis direction, and $W(x_j - X_i)$ represents the contribution weight of particle $j$ to grid point $i$, indicating the influence range of particle j's spatial distribution.

Conventional particle motion solvers may encounter numerical instability during the update process due to the influence of the magnetic field on the velocity. In contrast, the Buneman-Boris method addresses this issue by separately applying half-step and full-step updates for the electric and magnetic field effects, respectively. This approach allows for a more accurate reflection of the electromagnetic field's influence on the particle's velocity and position during the update process :

$$v_{n+1/2} = v_n + \frac{q_e}{m_e} E \frac{\Delta t}{2} \tag{21}$$

$$v_{n+1} = v_{n+1/2} + \frac{q_e}{m_e} v_{n+1/2} \times B \Delta t \tag{22}$$

$$r_{n+1} = r_n + v_{n+1} \Delta t. \tag{23}$$

Here, the symbol $r_n$ represents the particle position at the n-th time step, and $v_n$ represents the particle velocity at the n-th time step. The Buneman-Boris method primarily employs explicit numerical integration to solve the particle motion equations, making it suitable for high-precision electromagnetic particle simulations. In the study of two-stream instability, it enables a more accurate calculation of the interaction between particles and electromagnetic fields.

In plasma numerical simulations, the concept of the super-particle is introduced to simplify the computation and description of the system. The relationship between the super-particle and particle density can be expressed as:

$$n_i = \frac{N_p}{N_x \Delta x}. \tag{24}$$

$n_i$ represents the particle density. $N_p$ denotes the number of super-particles, and $N_x$ represents the total number of spatial grids.

In theory, as long as the grid is sufficiently fine and the particle number is large enough, the PIC simulation can be equivalent to solving the Vlasov equation, and in principle, it can compute the particle dynamics under most weak-collision or collisionless conditions. The FDTD method discretizes the electromagnetic field on a spacetime grid to calculate the propagation of the field. By combining the FDTD method with PIC simulations, the dynamic evolution of particles in the electromagnetic field can be effectively addressed [29-30]. The specific program-solving framework is shown in Figure 2.

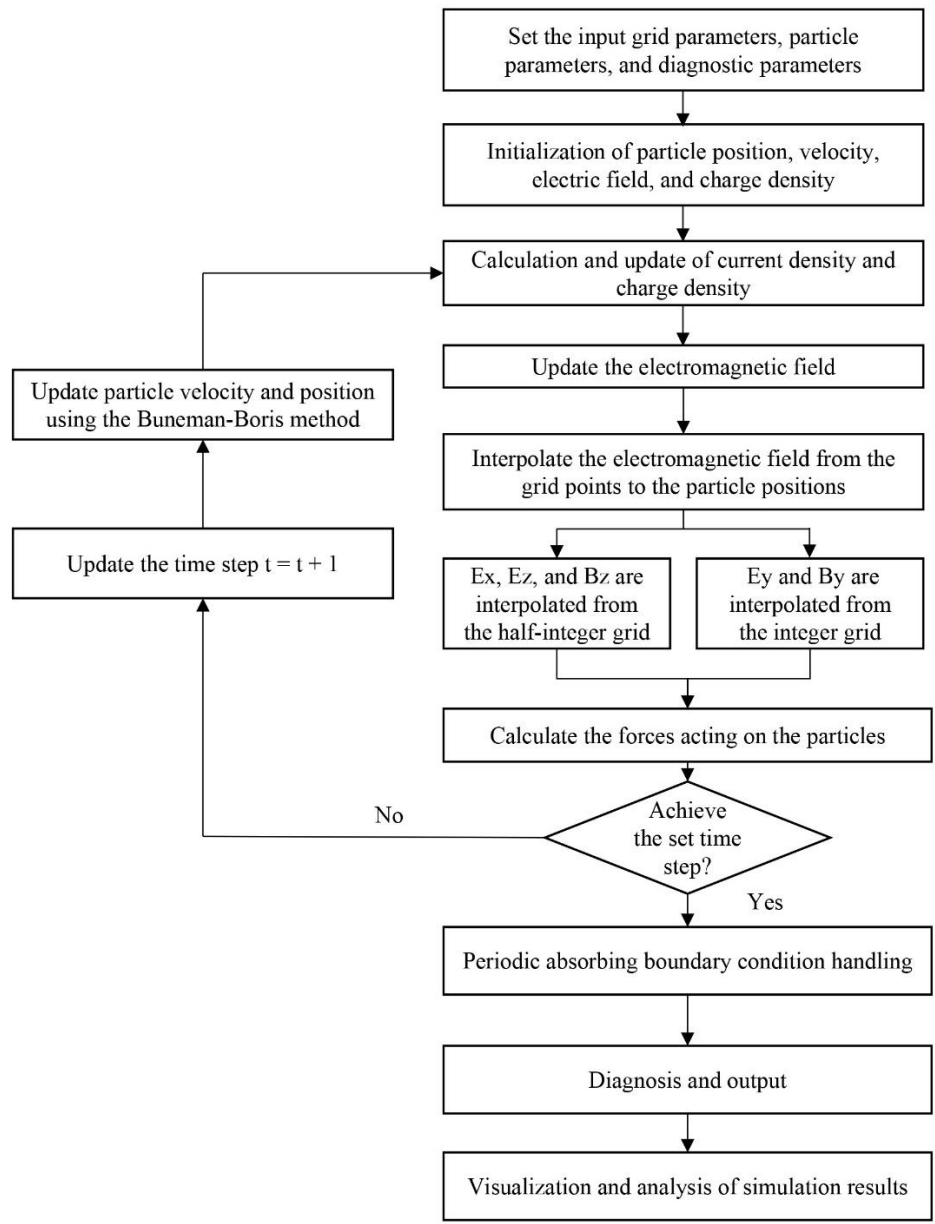

**Fig. 2.** The computational process of the Kyoto University Electromagnetic Particle Code: 1D version (KEMPO1) [28].

## 2.3 Inputs and conditions

The heating and acceleration of the solar wind continues to be one of the unresolved key problems in heliophysics. In typical solar storm events involving coronal mass ejections, the solar wind velocity significantly exceeds its normal speed. The Discover satellite, in near-Earth orbit, has recorded a maximum radial velocity of the solar wind reaching 820 km/s. Concurrently, the peak value of the southward magnetic field component is measured as 70 nT [31]. However, since the southward magnetic field is parallel to the electron motion velocity, we will focus on the impact of the radial magnetic field-induced electron cyclotron motion on two-stream instability in subsequent calculations. During a solar storm, the high-speed solar wind near 1 AU changes with radial distance, approximating an adiabatic process [32]. In the near-region, we assume no Alfvén wave energy heating during the limited time of solar wind electron motion. A negative correlation is observed between electron temperature and solar wind velocity. Based on previous studies from Parker Solar Probe (PSP) and WIND, we fitted the relationship between solar wind velocity and electron temperature near the cusp region [33], as shown in Figure 3. Cluster satellite observations show that the upward-moving electrons near the polar cusp are primarily from the ionosphere, with an upward electron flux greater than the downward flux [34]. Electron acceleration during disturbance events is a key feature. We assume that, during the brief occurrence of two-stream instability, the local electron density changes by 50%. Based on observational data from the Cluster satellites, we set the solar wind velocity in the polar cusp to be 700 km/s. The electron temperature is calculated to be $4.93 \times 10^4$ K based on the negative correlation between the solar wind velocity and electron temperature. During typical geomagnetic storms triggered by solar wind, the upward-moving electrons in the polar ionosphere are accelerated by strong parallel electric fields in the cusp region, ranging from 20-100 mV/m [35-36]. The drift velocity can reach 200 km/s, consistent with the acceleration range measured by Cluster. Given that electron acceleration is much more efficient than the increase in electron temperature, we set the electron temperature at 2500 K [37].

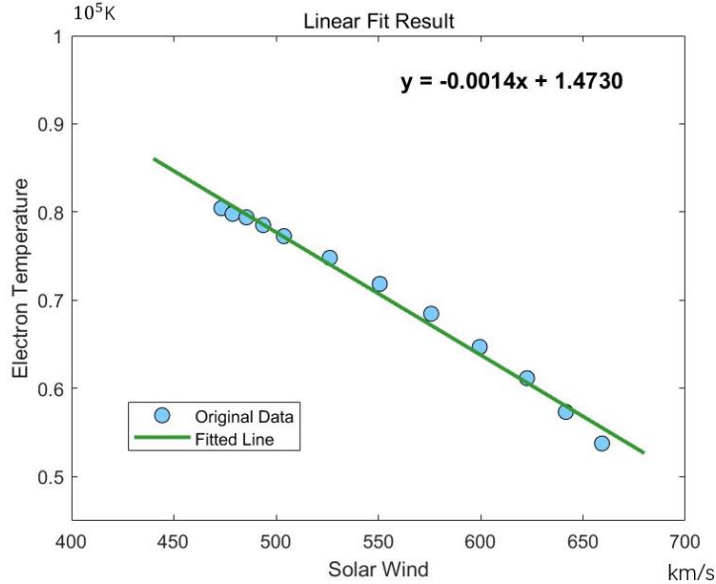

**Fig. 3.** Fitted curve of electron temperature and solar wind velocity in near earth space.

The goodness of fit, represented by an $R^2$ value of 0.9895 indicates that the fitted model explains 98.95% of the variance in the data. The linear fit effectively captures the negative correlation between solar wind velocity and electron temperature.

After determining the particle parameters, the PIC grid parameters are set in the KEMPO1 code as follows: DX=1.25, NX=128, DT=$4 \times 10^{-4}$, and NT=131072, where DX represents the grid spacing in the simulation, NX denotes the number of grid points, DT refers to the time step, and NT represents the number of time steps in the simulation run. To ensure the conservation of energy in the system, periodic boundary conditions are applied in the simulation. Particles that exit the simulation domain from one side will re-enter from the opposite side with the same velocity. The physical parameters in the simulation process are normalized to dimensionless forms. The purpose of normalization is to convert physical quantities into values independent of a specific unit system by selecting a set of reference quantities, thereby simplifying the numerical simulation process [38]. The reference quantities are based on two key physical constants: the speed of light, $c$ ($c_0 = 3000$), and the plasma frequency, $\omega_p$. The purpose of setting the speed of light $c$ to 3000 is to ensure that the normalized values of velocity and other physical quantities in the simulation remain within a range that is manageable and computationally efficient. Once these two quantities are determined, we can directly compute the normalized parameters, such as electric field strength, magnetic field strength, current density, and particle energy. The total plasma frequency $\omega_p^2 = \omega_{ps}^2 + \omega_{pi}^2$, where $\omega_{ps}$ and $\omega_{pi}$ represent the frequencies of solar wind and ionosphere electrons, respectively. The $\omega_p$ has been normalized, where $\omega_{ps}$ is proportional to $\sqrt{n_s}$ and $\omega_{pi}$ is proportional to $\sqrt{n_i}$. Therefore, the unit electric field and energy can be calculated as:

$$[E] = \frac{c}{c_0} \times \frac{\omega_p}{2\pi} \times \left|\frac{m}{q}\right| \tag{25}$$

$$[Energy] = \left(\frac{c}{c_0} \times \frac{\omega_p}{2\pi} \times \left|\frac{m}{q}\right|\right)^2. \tag{26}$$

## 3. RESULTS AND ANALYSES

In this section, we analyze the nonlinear evolution of the two-stream instability based on numerical simulation results. In the simulation, we focus on the two-stream instability between two electron beams, treating positive ions and protons as a uniform background. The interactions of the electrons are represented in the Cartesian coordinate system. Section 3.1 presents the evolution of electrostatic two-stream instability under solar storm perturbations. Section 3.2 examines the influence of spatially varying magnetic fields on the development of the instability. Section 3.3 provides an assessment of the potential adverse impacts of the two-stream instability on spacecraft operations.

### 3.1 Two-stream instability

The phenomenon of two-stream instability in the polar cusp may occur during the acceleration of upward-moving ionosphere electrons by strong electric fields in high-speed solar wind and geomagnetic storm conditions. Research has already confirmed the existence of ESWs in the polar cusp [39]. Observations from the Cluster satellites indicate that in the polar cusp, particularly in the region where the magnetopause intersects the magnetosphere boundary, the energy injection into the Earth's magnetosphere caused by solar storms is often accompanied by anisotropy in the electron thermal velocity. Due to the effects of magnetic field confinement and plasma dynamical processes during the propagation of the solar wind, solar wind electrons often exhibit anisotropic distributions, with more pronounced anisotropy in high-speed solar wind. Heyu Sun investigated the instabilities driven by solar wind electrons under different electron thermal velocity distributions, and found that the ratio $T_\perp/T_\parallel$ generally appears more frequently within the range of 0.5–2 [40]. To validate the potential excitation and specific manifestations of the two-stream instability, we first selected solar wind electron beams from the Northern Hemisphere with $T_\perp/T_\parallel=2$ as the subject for our study of two-stream instability.

The two-stream instability is a type of electrostatic instability, and an electrostatic model is employed in Section 3.1 for simulation. According to equation $v_{th} = \sqrt{2k_B T/m_e}$, the horizontal and vertical thermal velocities of electrons are calculated to be 669.4 km/s and 946.8 km/s, respectively. The plasma frequency is normalized, with the plasma frequencies of solar wind electrons and upward-moving ionosphere electrons denoted as $\omega_{ps} = 0.586$ and $\omega_{pi} = 0.817$. Figure 4 illustrates the velocity variations of electrons in the phase space when the two-stream instability is excited.

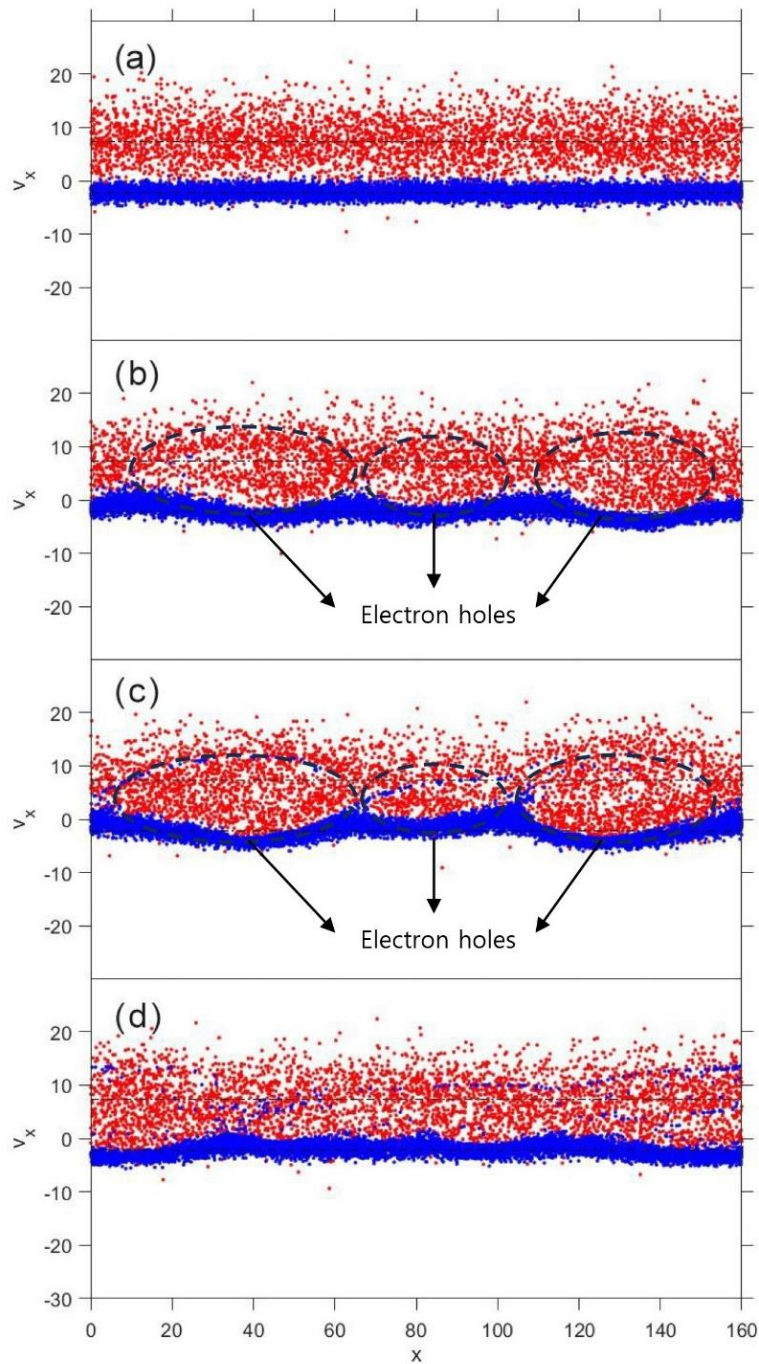

**Fig. 4.** The phase space evolution of the electron velocity distribution of the two-stream instability at different times: (a) $\omega_p t=0$, (b) $\omega_p t=20.28$, (c) $\omega_p t=32.27$, (d) $\omega_p t=52.43$.

In Figure 4, four time intervals are selected for analysis, which show significant changes in the particle motion states in phase space. The horizontal axis $x$ denotes the normalized spatial position and the vertical axis $v_x$ represents the normalized electron velocity. At the initial moment, the solar wind electron beam and the upwardly accelerated ionosphere electrons both exhibit no discernible fluctuations, and the interactions between the electrons are not significant. At $\omega_p t=20.28$, the electron velocity distribution in phase space exhibits small-amplitude oscillations, and the instability begins to grow. Electron holes start to form between the solar wind electron perturbations and the upward-moving ionosphere electrons, but these structures are not very distinct and are widely dispersed, indicating that the instability is still in its early stage of development. At $\omega_p t=32.27$, the electron velocities in phase space begin to exhibit pronounced fluctuations, with the instability gradually increased and distinct vortex-like electron hole structures emerged. The formation and evolution of electron hole structures are related to the nonlinear effects of the electric field.

During this process, the velocity distribution layering structure of the two electron beams in phase space gradually becomes indistinct. Finally, at $\omega_p t$=52.43, the potential wells of adjacent cavities begin to attract each other, causing the electron-hole structure of the dual-stream system to gradually merge. At this stage, larger electron-hole pairs do not form. Instead, due to electrostatic fluctuations and Landau damping, energy is likely transferred back to the electrons, whose velocity differences have previously diminished, resulting in the randomization of electron velocities and the gradual blurring of the electron-hole boundaries.

From the perspective of the static BGK solution of the Vlasov-Poisson equation, the formation and merging evolution of electron holes result from the self-consistency of the particle distribution function and the nonlinear effects of the electric field. The electric field is closely linked to the changes in the particle distribution, and the electric field induced by the perturbation further affects the particle trajectories. The particle distribution function described by the BGK solution tends toward a new equilibrium point, which may differ from the initial equilibrium. At this stage, the electron holes could not completely dissipate but rather persist in some form within the system. These holes could be re-excited or reconfigured by new perturbations. Electron holes in phase space correspond to ESWs, and the evolution of ESW structures is illustrated in Figure 5.

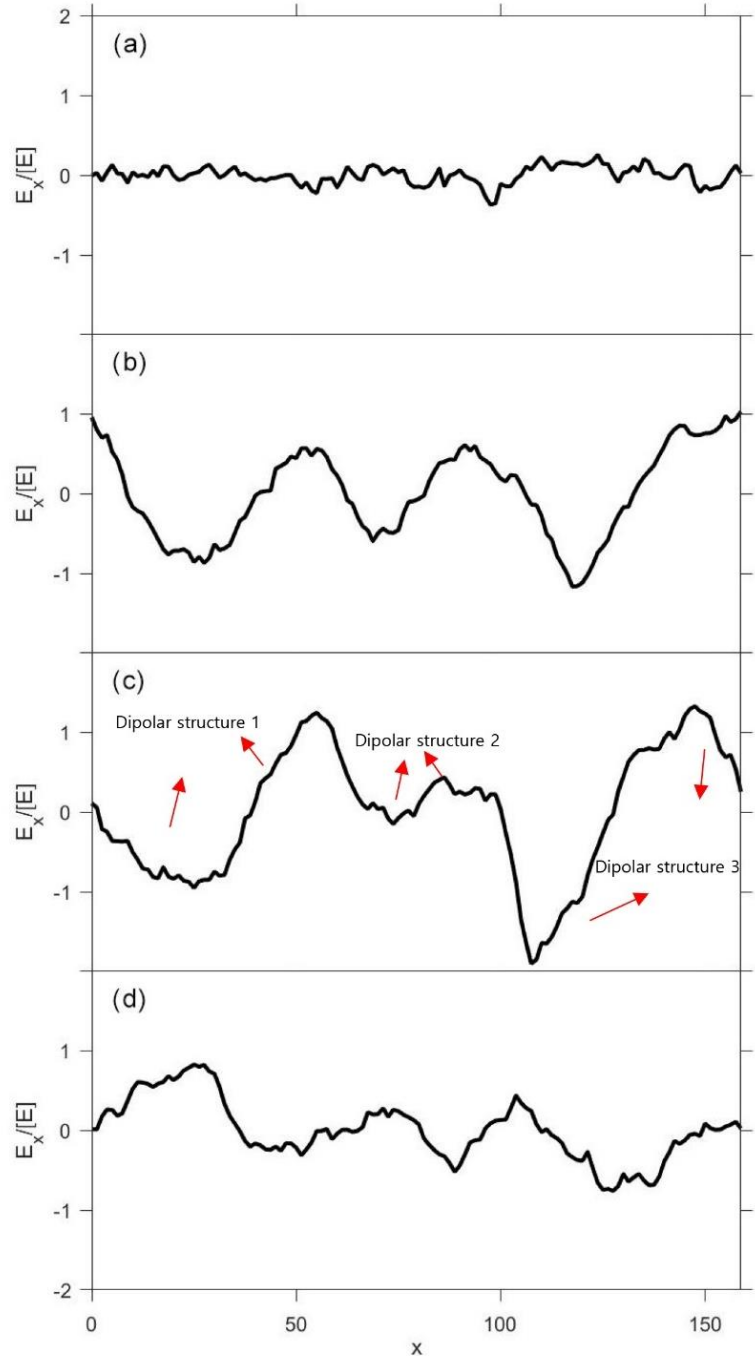

**Fig. 5.** The electric field fluctuation excited by the two-stream instability along the $E_x$ direction at different times: (a) $\omega_p t=0$, (b) $\omega_p t=20.28$, (c) $\omega_p t=32.27$, (d) $\omega_p t=52.43$.

In the actual space environment, these fluctuations can be interpreted as ESWs. ESWs are primarily generated through the nonlinear evolution of the two-stream instability and correspond to electron holes in phase space. ESWs typically exhibit a dipolar structure, reflecting the trapping (acceleration) and reflection (deceleration) effects of potential wells on electrons, with a typical duration ranging from 0.1 to 10 ms [41]. At $\omega_p t=32.27$, three vortex structures in phase space and three distinct dipolar electric field fluctuations are observed, with the coordinates of the electric field fluctuations aligning with those of the electron holes. The growth of ESWs intensifies the interactions between electron beams, thereby enhancing the instability of the two-stream system. When electron vortex structures emerge, a large number of electrons are concentrated in localized regions, forming high-density electron clusters. These clusters generate strong induced electric fields under the influence of the electric field, which significantly amplifies the amplitude of ESWs. Therefore, ESWs and the two-stream instability promote each other and are positively correlated. As the instability progresses, the amplitude of the electric field fluctuations increases, and the interactions between the electron populations become more pronounced. However, the electric field fluctuations do not grow indefinitely. When the electron velocity approaches the phase velocity of the electric field wave, electrons resonate with the wave, leading to the reabsorption of the wave's energy. At this point, Landau damping exerts a significant weakening effect on the electric field fluctuations, causing the energy of the electric field waves to gradually decrease.

As shown in Figure 6, the initial bimodal distribution of electron velocities arises from the drift velocity and thermal velocity specified by Equation (1). At the initial moment, the blue dashed curve exhibits a tail enhancement, which exhibits an energetic tail. As the two-stream instability evolves, the electron velocity distribution is governed by the Vlasov-Poisson equations. The energetic tail gradually diminishes, accompanied by a decrease in the number of tail electrons. The interaction between electrons and electric field fluctuations leads to energy redistribution, causing the tail electrons to progressively lose energy and migrate toward the lower-velocity region.

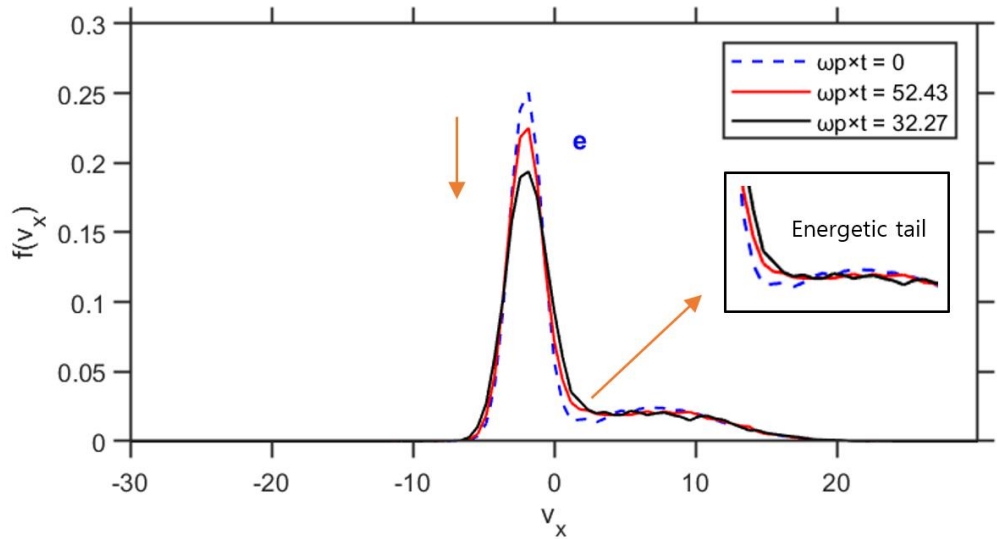

**Fig. 6.** The electron velocity distribution function of two-stream instability.

At $\omega_p t = 32.27$, distinct phase-space electron holes are formed due to the two-stream instability, leading to a more uniform particle distribution. By $\omega_p t = 52.43$, the ESWs gradually attenuate, the two-stream instability subsides, and the proportion of low-velocity electrons increases, resulting in an approximate single-peak distribution. Recent research findings indicate that, despite ESWs being a form of electric field fluctuation, they primarily serve as a medium for energy exchange between electrons [42]. The electron velocity distribution does not result in a significant shift in the overall velocity distribution interval. A more rational explanation for why ESWs do not accelerate electrons is that they affect electrons through localized potential disturbances. The upstream and downstream regions in phase space exhibit opposing electric field distributions, causing the influence of the wave on electrons to manifest primarily as local changes in velocity rather than sustained acceleration.

The two-stream instability predominantly manifests as an electrostatic two-stream instability, since its driving mechanism arises directly from the relative drift velocity between electron beams. In contrast, the development of electromagnetic two-stream instability requires much stronger magnetic confinement or relativistic conditions to become significant, and thus it does not dominate under most plasma environments. Figure 7 illustrates the variation of energy components in the two-stream system.

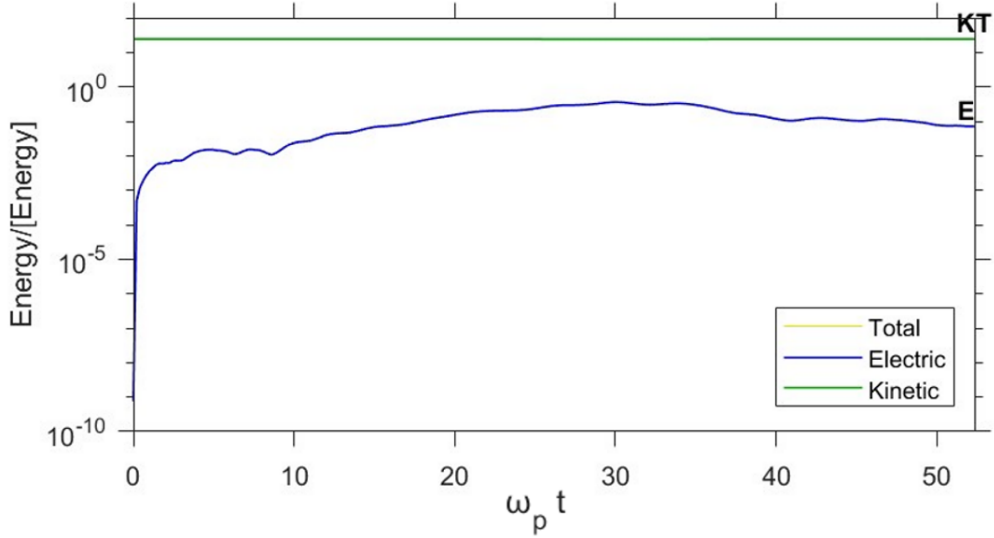

**Fig. 7.** The changes in total energy, electric field energy and kinetic energy during the evolution of two-stream instability.

In the electrostatic mode, the energy conversion in the two-stream instability primarily involves the kinetic energy and the electrostatic potential energy, as indicated by Equation (11). Kinetic energy dominates the total energy of the system, with the relative motion, oscillation, and acceleration of the solar wind electron beams and the upwardly accelerated ionosphere electrons serving as the primary sources of energy. Initially, the electric field energy increases rapidly, primarily due to the conversion of the beam's kinetic energy into electrostatic wave energy through Landau resonance. As the instability approaches saturation, the boundaries of the phase space cavities become blurred due to electron escape, and the structure gradually disintegrates. The localized ESWs of the electron holes are progressively weakened, potentially transitioning toward a more isotropic Maxwellian distribution. In terms of energy density, the electric field energy remains several orders of magnitude smaller than the kinetic energy.

The energy distribution reveals the conversion of electron kinetic energy into electric field energy and its evolutionary process. The dispersion relation determines the growth rate of waves and the evolutionary characteristics of the instability. In Equation (9), the Fried–Conte function $Z(\zeta_j)$ describes the interaction between electrons and electric field fluctuations, particularly the Landau resonance effect, which makes the established dispersion relation more representative of the actual spatial context. The dispersion relation shown in Figure 8 reflects the enhancement and attenuation of $E_x$ along the electron motion direction in terms of frequency and wave number.

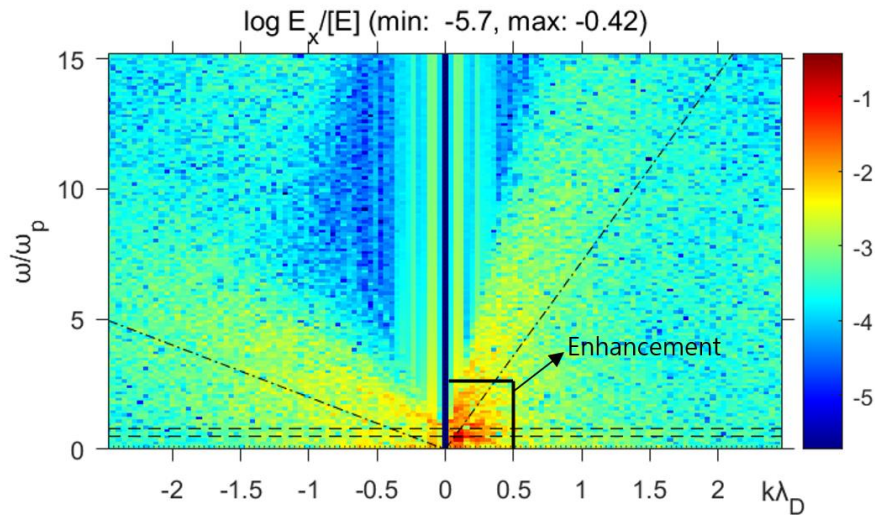

**Fig. 8.** The dispersion relation of $E_x$ in the electrostatic mode for two-stream instability.

In the theoretical analysis, the expression of the Fried–Conte dispersion function was introduced, but the real part of the expansion in Equation (7) was not described in detail. Since the phase velocity of two-stream instability waves in space plasma is close to the electron drift velocity, the numerator of $\zeta_j$ in Equation (6) and the distribution can be regarded as being of the same order of magnitude. Under this condition, a power series expansion can be applied, and the real part of the function $Z(\zeta_j)$ can be expressed as follows:

$$Re[Z(\zeta_j)] = -2\left(\zeta_j + \frac{2}{3}\zeta_j^3 + \frac{4}{15}\zeta_j^5 + \cdots\right). \tag{27}$$

Furthermore, the complete expression of the $Z(\zeta_j)$ function can be obtained as follows:

$$Z(\zeta_j) = i\sqrt{\pi}e^{-\zeta_j^2} - 2\left(\zeta_j + \frac{2}{3}\zeta_j^3 + \frac{4}{15}\zeta_j^5 + \cdots\right). \tag{28}$$

Electrostatic fluctuations driven by two-stream instability are predominantly distributed in the low-wave-number regime, where ESWs are markedly amplified. From a frequency perspective, these fluctuations are mainly concentrated in the low-frequency range close to the plasma frequency. According to Equation (28), when the wave frequency approaches the plasma frequency, $\zeta_j \to 0$, where the imaginary part exerts the strongest influence on the dispersion relation. The imaginary part corresponds to the Landau resonance term and reflects the intensity of energy exchange between electrons and waves. As the imaginary part increases, the interaction between resonant electrons and the wave becomes most pronounced, which is manifested in the dispersion relation diagram as a deepening of color. The real part of the dispersion relation, after neglecting higher-order infinitesimal terms, can be approximated as $-2\zeta_j$, exhibiting a linear dependence. This indicates that the amplitude of ESWs becomes more sensitive to frequency variations, which is reflected in the dispersion relation diagram by a more distinct gradient of the electric field near the plasma frequency. In addition, the pronounced enhancement of ESWs is primarily concentrated in the positive half-axis range of normalized $k\lambda_D$ from 0 to 0.5, which corresponds to the motion of solar wind electrons; therefore, solar wind electrons serve as the primary excitation source of ESWs.

## 3.2 Effects with magnetic field

The analysis of the two-stream instability in the polar cusp region under the electrostatic mode primarily focuses on the excitation of electrostatic waves between electron beams and the evolution of electron phase space. Due to the presence of the magnetic field, the anisotropy of electrons and the electron cyclotron frequency significantly influence their motions. In Section 3.2, we investigate the impact of factors such as the angle between the magnetic field and the wave vector, as well as electron thermal velocity anisotropy, on the evolution of two-stream instability and ESWs under the electromagnetic mode. The schematic diagram of electron motion with respect to the magnetic field in the polar cusp is shown in Figure 9.

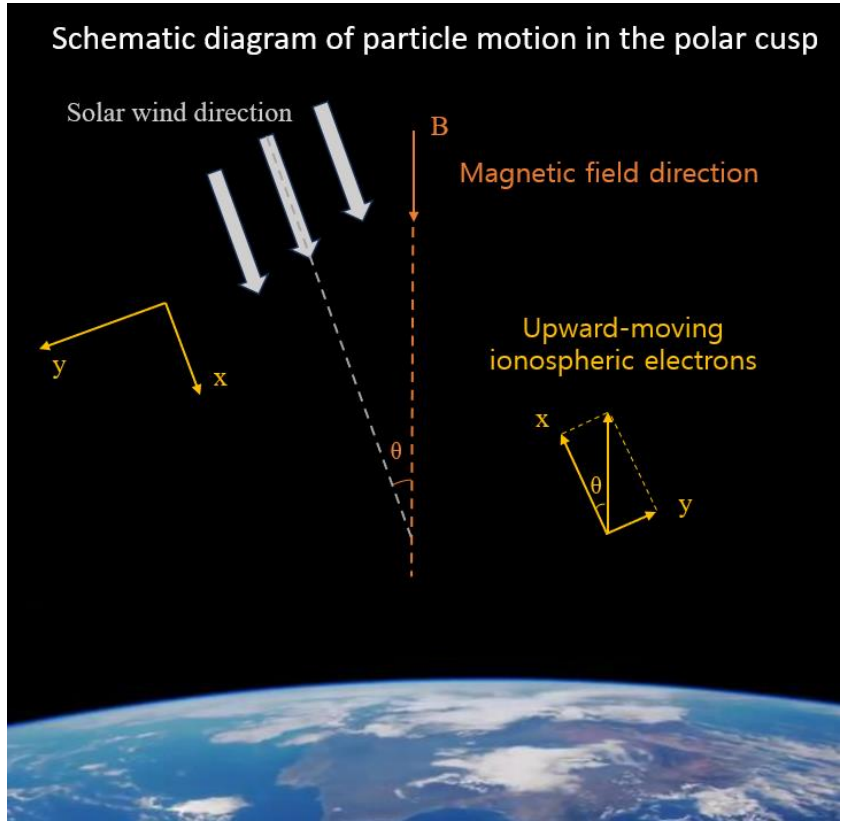

**Fig. 9.** Schematic diagram of electron motion considering the magnetic field and *θ* in the polar cusp.

In the context of the polar cusp region, we introduce a dimensionless parameter $\omega_c/\omega_p$, which is defined as Equation (29). According to previous study, the magnetic field strength in the polar cusp is typically low, and the plasma density is relatively high. We set $\omega_c/\omega_p = 0.1$ to ensure that the simulation results are consistent with the plasma behavior in the polar cusp [43]. When $\omega_c/\omega_p < 1$, wave and plasma effects dominate, with electron cyclotron motion being considered as a secondary factor.

$$\frac{\omega_c}{\omega_p} = \frac{B}{2\pi}\sqrt{\frac{\varepsilon_0 m_e}{n_e e^2}} \tag{29}$$

The polar cusp region is a critical area for the coupling of solar wind, magnetosphere, and ionosphere. Typically, the angle *θ* between the magnetic field and the wave vector in the polar cusp region is close to 0°, and the ESWs excited by two-stream instability propagate approximately along the magnetic field direction. The fluctuations generated by instability propagate along the magnetic field direction. In Section 3.1, the excitation of two-stream instability for the case of *θ*=0° is also analyzed. In the actual polar cusp region, *θ* is driven by magnetic reconnection and solar storm factors, and can range from 0° to 45°, though statistically, *θ*<30° predominates [44]. In the simulation, the electrostatic solitary wave fluctuation $E_x$ propagates along the x-axis. We consider the angle between the magnetic field and the wave vector as the angle between the magnetic field and the x-axis direction. The evolution of fluctuations at different angles over time and space is shown in Figure 10.

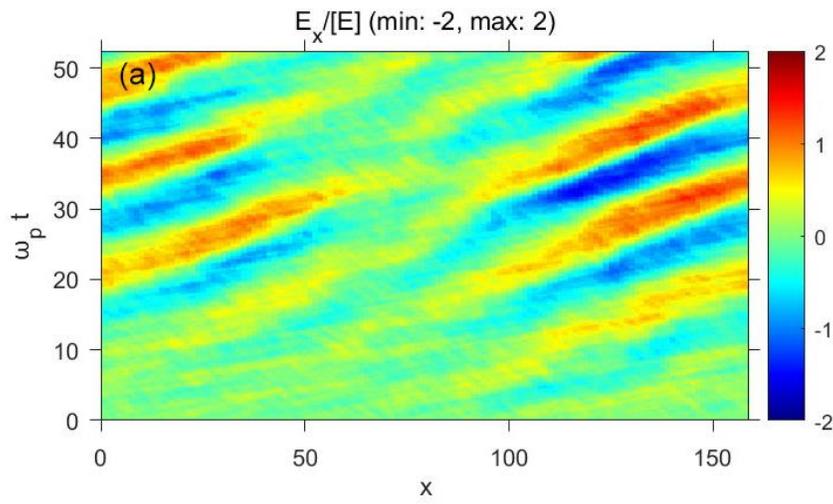

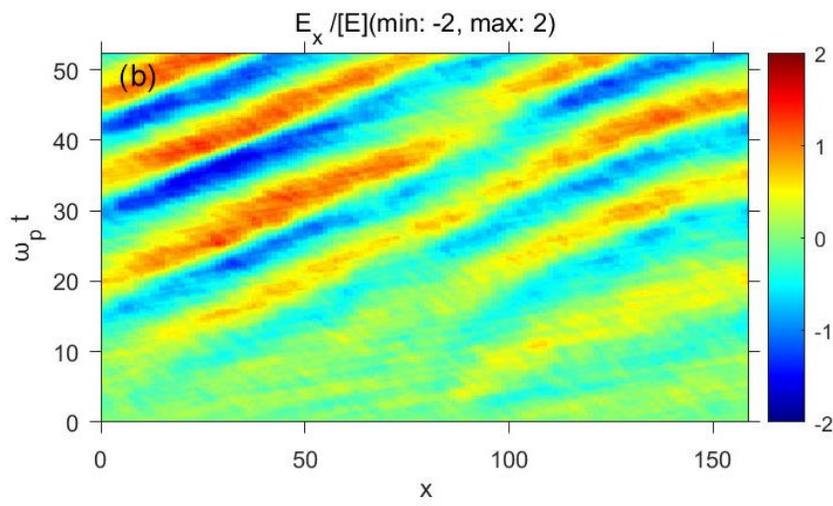

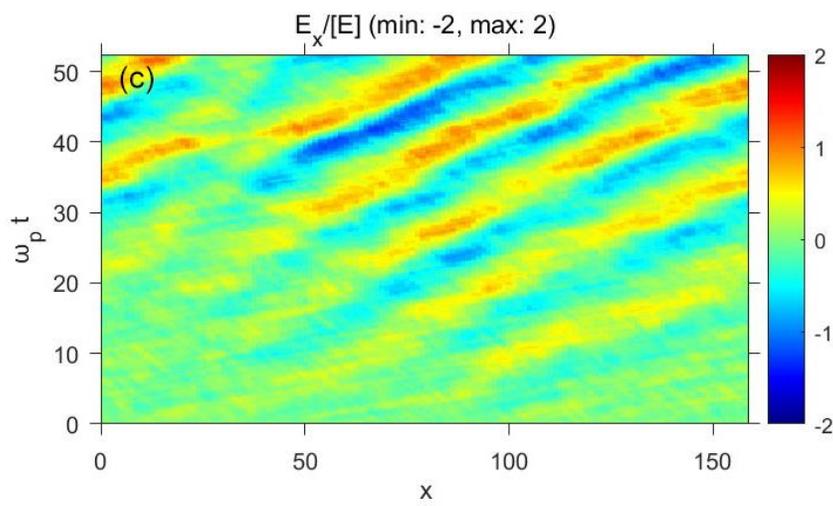

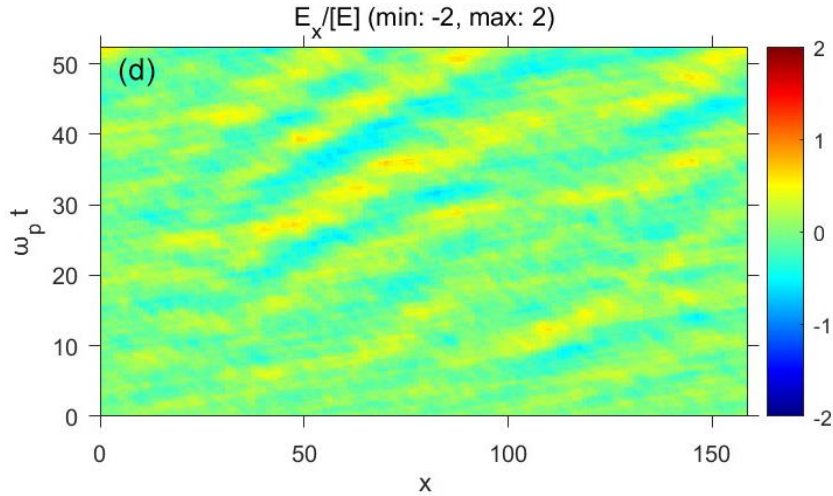

**Fig. 10**. The oscillation behavior of the electric field along the $E_x$ direction under different angles between magnetic field and wave vector: (a) $\theta=0°$, (b) $\theta=15°$, (c) $\theta=30°$, (d) $\theta=45°$.

In magnetized plasmas, the two-stream instability in electromagnetic modes experiences competing effects from Landau damping and electron cyclotron damping. During the growth phase of the instability, beam electrons become trapped within evolving electron holes, continuously injecting energy into ESWs. Concurrently, Landau damping dissipates wave energy, causing ESWs attenuation. At parallel propagation ($\theta=0°$), Landau damping reaches maximum efficiency. Increasing the angle $\theta$ reduces Landau damping rate while enhancing electron cyclotron damping rate. The weakened Landau damping fails to compensate for intensified cyclotron damping. Beyond $\theta>30°$, both the instability growth and ESWs amplification become suppressed. Electron cyclotron motion converts wave energy parallel to particle motion into perpendicular thermal energy. Simultaneously, the beam electron velocity along the x-axis drops below the parallel phase velocity of the electrostatic waves. This velocity mismatch significantly reduces the resonant electron population, thereby depriving the ESWs of sustained energy supply.

Figure 10b demonstrates maximum bipolar amplitude of ESWs at $\theta=15°$, identifying this angle as the optimal damping equilibrium where the competition between Landau damping and electron cyclotron damping is minimized. The electron interactions along the x-axis result in insufficient Landau damping, while moderate $\theta$ values suppress electron cyclotron damping. Consequently, peak electric field amplitudes emerge. Within $0°<\theta<30°$, distinct ESWs propagate along the $E_x$ direction. At $\theta=45°$, however, electron cyclotron damping is the primary driver of energy dissipation. This disrupts velocity resonance matching between waves and particles, consequently suppressing two-stream instability development.

We have extracted images of the electrostatic wave fluctuations in the $E_x$ direction at different values of $\theta$ when the two-stream instability growth is near saturation at $\omega_p t=32$, as shown in Figure 11. The amplitude of $E_x$ electrostatic fluctuations increases by 48.7% at 15° compared to that at 0°, while it decreases by 28.6% at 45° relative to the amplitude at 0°. This suggests that the two-stream instability is more pronounced near 15°, where the larger amplitude of ESWs results in a more significant accumulation of electrons. Interactions between solar wind electrons and upwardly accelerated ionosphere electrons result in rapid growth of the instability. Above 45°, the electrostatic fluctuations exhibit significant attenuation, accompanied by blurred phase-space vortices and enhanced electron thermal mixing, which effectively suppress the two-stream instability and its adverse effects on spacecraft and satellite systems.

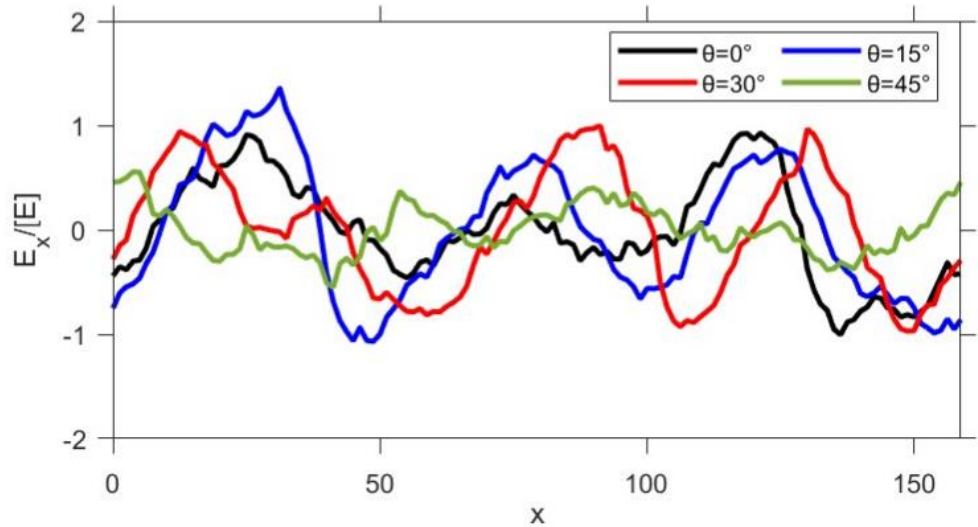

**Fig. 11**. The evolution of electrostatic fluctuations at different values of $\theta$.

The magnetic field causes the electron motion to become anisotropic, so we further analyzed the impact of electron temperature anisotropy on instability under electromagnetic modes. When the electron temperature is maintained constant, the electron velocity distribution function transitions from a bimodal to a unimodal structure if the velocity distribution is isotropic or if the horizontal thermal velocity is dominant. A broader electron distribution function enhances Landau damping, preventing the beam electrons from transferring energy to ESWs. In the unimodal distribution, where the electron thermal velocity is dominant, electron trajectories tend to be more diffusive. Under these conditions, the two-stream instability between two electron beams can be suppressed. Comparing the results in Figure 12 with those in Figures 10(a), 10(b), and 10(c), it is evident that the amplitude of ESWs excited along the direction of electron motion is reduced. The distinct bipolar electric field distribution between adjacent positions is no longer observed. Therefore, the anisotropic thermal velocity of electrons effects the growth of the two-stream instability.

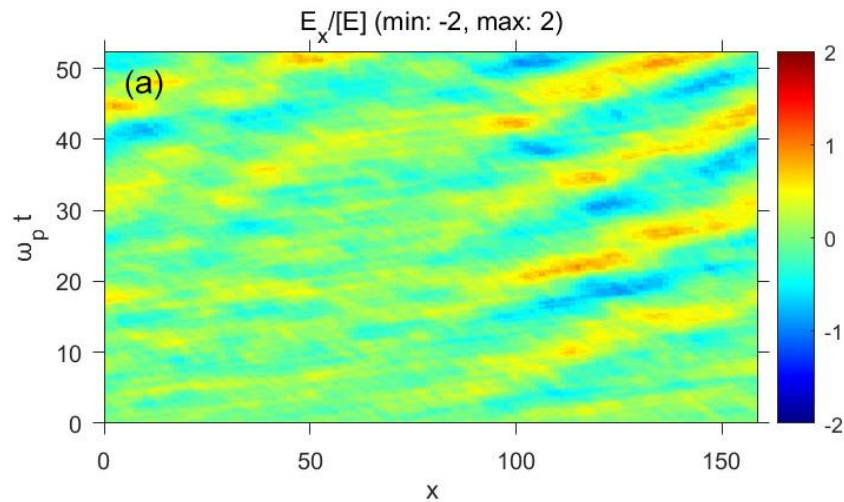

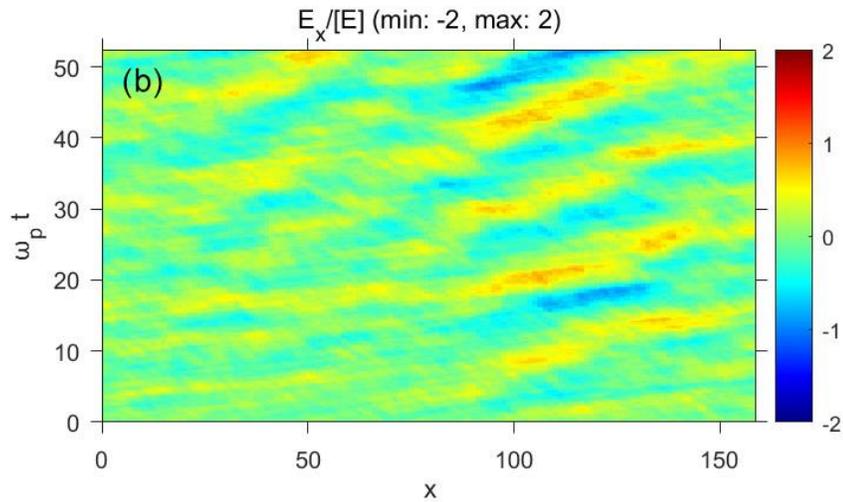

**Fig. 12.** Evolution of $E_x$ fluctuations over time and space: (a) Isotropic distribution of electron thermal velocities in both horizontal and vertical directions; (b) Anisotropic distribution with dominant horizontal electron thermal velocities, where $T_\perp/T_\parallel=0.5$.

  The increase in the thermal velocity of electrons in the horizontal direction makes it difficult for the potential well to capture enough electrons to form a vortex structure in phase space. The BGK mode, as a nonlinear steady-state solution, needs to overcome a certain degree of Landau damping. The single-peak structure of the electron distribution function disrupts the conditions necessary for the existence of the BGK mode. Enhanced Landau damping dissipates the initial perturbation before it has a chance to develop into a nonlinear BGK structure. This fundamentally suppresses the excitation and maintenance of the BGK mode, making it difficult to form a stable electron hole structure. The electric field fluctuations observed in Figure 12 are mainly caused by the relative motion of the solar wind electron beam and the upward accelerated ionospheric electron beam, which leads to random fluctuations in local electron density, as shown in Figure 13.

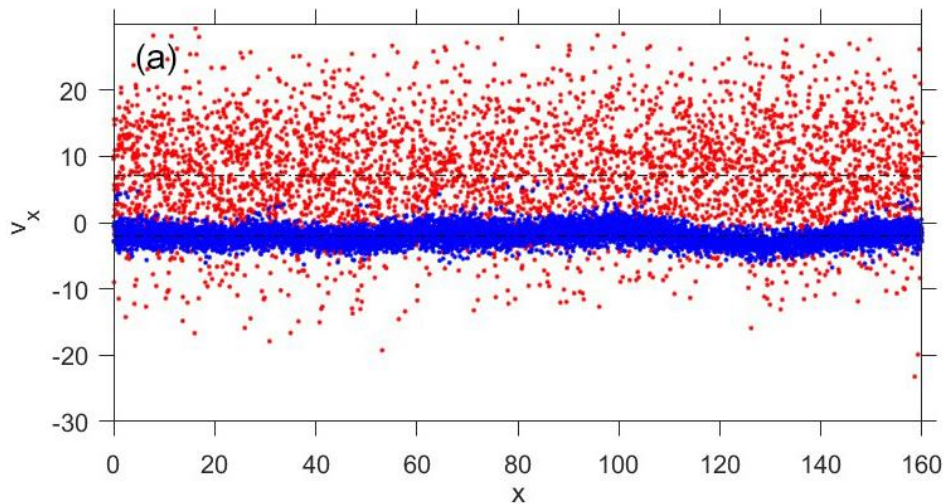

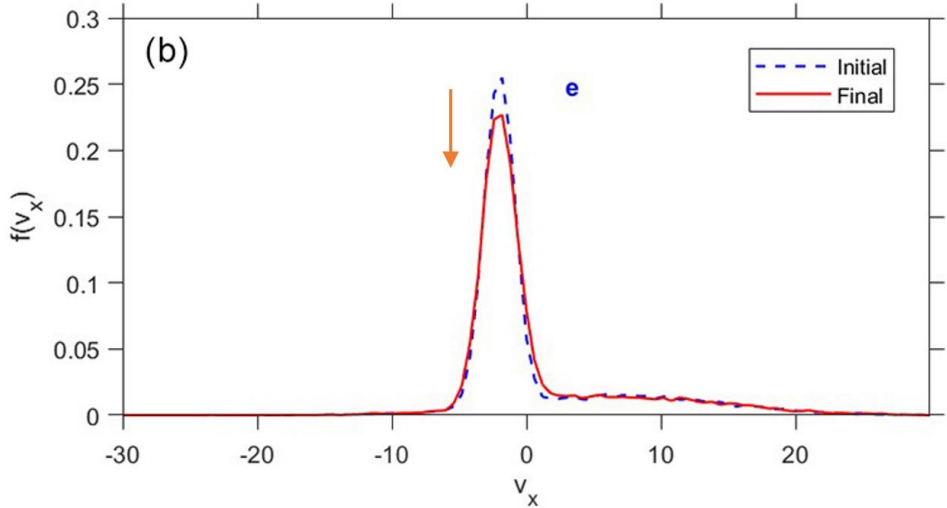

**Fig. 13**. Characteristics of electrons with isotropic thermal velocity distribution at time $\omega_p t$ = 52.43: (a) electron velocity distribution in phase space, (b) single-peak electron velocity distribution function.

Based on the simulation discussions in sections 3.1 and 3.2, we summarize the possible conditions for the excitation of ESWs and two-stream instability by solar wind electrons in the polar cusp and upwardly accelerated ionosphere electrons. The excitation of two-stream instability is particularly favored when the electron drift velocity and the thermal velocity in the drift direction are comparable or when the drift velocity is slightly greater than the thermal velocity. During periods of solar storms, the solar wind often exhibits significantly higher drift velocities. Specifically, when the instantaneous drift velocity reaches above 700 km/s, the corresponding electron temperature typically falls below 50000 K, as shown in Figure 3, thereby facilitating the excitation of two-stream instability in the polar cusp region. Two-stream instability is more likely to be triggered in a spatial environment characterized by an anisotropic distribution of electron thermal velocities, with a dominant perpendicular thermal velocity component. Under such conditions, the electron thermal velocity along the drift direction is relatively low. The inherent velocity difference in the bi-Maxwellian electron distribution function facilitates fluctuations and electron energy exchange, thereby providing favorable conditions for the excitation of two-stream instability. In the presence of a magnetic field, the excitation of two-stream instability is most probable when the angle $\theta$ between the magnetic field and the drift velocity is within the range of 0°–30°. Both Landau damping and electron cyclotron damping play significant roles in modulating the growth of two-stream instability. When $\theta$ exceeds 45°, two-stream instability and ESWs are effectively suppressed. Conversely, when the electron thermal velocity distribution is isotropic or when the parallel thermal velocity component is dominant, the monomaxwellian electron distribution function results in strong Landau damping, which continuously dissipates the energy of electrostatic fluctuations, thereby inhibiting the development of two-stream instability.

### 3.3 The impact on spacecraft

Currently, the study of plasma instability and its impact on spacecraft operational stability is receiving increasing attention [45]. Based on the aforementioned study of two-stream instability, this paper develops a model that describes the phenomena of two-stream instability which may cause surface charge accumulation and communication frequency band of spacecraft. In addition to the supplementary current $I_{TSI}$ generated by two-stream instability, the spacecraft surface also experiences electron current $I_e$, ion current $I_i$, and electron secondary emission current $I_{sec}$. $V_s$ is used to represent the charging voltage on the spacecraft's surface. Therefore, the spacecraft surface charging process can be described by the following equation:

$$\frac{dV_s}{dt} = \frac{1}{C}(I_e + I_i + I_{sec} + I_{TSI}). \tag{30}$$

The capacitance C of the spacecraft surface determines its ability to store surface charge. The Equation (30) comprehensively accounts for the contributions of electron and ion currents from the surrounding plasma, secondary electron emission effects, and additional currents. The expression for the additional current $I_{TSI}$ generated by two-stream instability is given as follows:

$$I_{TSI} = en_e v_e \delta E_0 \cdot A. \tag{31}$$

In this equation, $v_e$ represents the average velocity that influences the movement of surface charges on the spacecraft, while $A$ denotes the surface area of the spacecraft that is affected by two-stream instability. The accumulation of surface charge is induced by the electric field perturbation $\delta E_0$ generated by the two-stream instability. The primary source of this electric field energy is the dissipation of electron kinetic energy, which is converted into electric field fluctuations. In practical estimates, we can substitute the amplitude of the electrostatic solitary wave into the calculation of $\delta E_0$. According to Equations (30) and (31), the electric field fluctuations induced by the instability lead to variations in the spacecraft surface potential. The bipolar distribution of ESWs further alters the motion of electrons. The rate of charge accumulation on the local surface area increases with the rise in potential difference.

Previous studies have indicated that electron holes in phase space correspond to local potential wells and continuously capture electrons to maintain their structure. The electrons carried by electron holes are highly susceptible to impacting spacecraft surfaces, thereby inducing non-uniform charging and leading to electrostatic discharge events. Such events may directly damage sensitive electronic components or couple into the electrical circuitry, generating electromagnetic noise. ESWs manifest as isolated potential structures, with their frequency spectrum concentrated in the region near or below the plasma frequency. When the energy of these noise components falls within the sensitive frequency bands of the spacecraft system, it can lead to signal distortion in electronic devices and a degradation of operational stability [46]. ESWs form predominantly in regions where the angle between the wave vector and the magnetic field is less than 30°. Within this range, electrons experience weaker magnetic confinement, making it easier for instabilities to be excited.

To mitigate the adverse effects of the instability on spacecraft, the following preventive strategies could be implemented: when a satellite enters high-risk regions such as the polar cusp and magnetosheath, it is essential to employ electric field detectors to monitor the amplitude of ESWs in real time. Moreover, the satellite orbit inclination should be designed to avoid regions with strong plasma turbulence whenever it is possible [47]. The electron density impacting the spacecraft surface along the magnetic field lines is higher than other direction, which increases the risk of discharge. Therefore, when operating in the polar cusp region, the spacecraft should prioritize avoiding the magnetic field direction aligned with the solar wind injection direction. Furthermore, materials with high secondary electron emission, such as indium tin oxide coatings, should be selected to trigger net charge emission upon electron incidence, which could thereby suppress the accumulation of negative potential [48].

## 4. CONCLUSION AND REMARKS

This study analyzes the evolution of two-stream instability in the polar cusp region under solar storm disturbances. By integrating existing literature and satellite-observed solar storm and electron motion data, we discuss the evolution process, dispersion relations, and potential negative impacts of two-stream instability on spacecraft. Key findings obtained through numerical simulation and theoretical analyses are summarized below:

1) Under solar storm disturbances, the interaction between solar wind electrons and upwardly accelerated ionosphere electrons within the plasma background can excite two-stream instability. The ESWs induced by the instability exhibit dipolar structures and amplitude variations. The formation of electron holes enhances the amplitude of ESWs and the accumulation of electric field energy. These ESWs also facilitate the formation of vortex-like electron hole structures. Electrons are trapped in the potential wells and undergo localized oscillations rather than continuous acceleration; thus, ESWs do not directly accelerate charged particles.
2) Two-stream instability typically exhibits a strong response in the low-frequency range and near the plasma frequency. As the frequency increases (in the high wave number region), the intensity of the waves gradually decreases. High-frequency fluctuations are generally suppressed by strong Landau damping effects. Although high-frequency fluctuations can occur in two-stream instability, their energy propagation is limited due to their shorter wavelengths. The waves excited by the two-stream instability are typically concentrated in the low-frequency range, predominantly in the form of low-frequency electrostatic waves propagating along the direction of particle motion.
3) During the interaction of electron beams, the angle between the electrostatic wave vector and the magnetic field effects the evolution of the two-stream instability. When the angle approaches 15°, the suppressive effects of Landau damping and electron cyclotron damping are weakest, leading to the excitation of a strong two-stream instability. For angles greater than 45°, electron cyclotron damping suppresses the growth of the two-stream instability. When the horizontal thermal velocity of electrons dominates or results in a single-peak distribution function, the enhanced Landau damping suppresses the excitation and maintenance of BGK modes, and the two-stream instability is also inhibited.

4) The fluctuations induced by the two-stream instability can increase the accumulation of surface charge on spacecraft and generate electromagnetic noise, which may cause damage to both surface materials and electronic systems. In practical spacecraft operations, it is crucial to orient sensitive surfaces to avoid alignment with the magnetic field and the direction of solar wind injection, while also monitoring the amplitude of ESWs in real time. Future research should focus on effectively monitoring the risks associated with extreme space weather and plasma instabilities. It is of great importance to enhance the survivability and mission reliability of spacecraft operating in the complex plasma environment of the polar cusp.

Currently, no experiment is available to demonstrate that the two-stream instability and ESWs cause negative interference to spacecraft in orbit operations. The theoretical analysis and simulation results in this paper reveal that the two-stream instability can potentially lead to spacecraft surface charging, electron accumulation, and electromagnetic interference with onboard equipment. In the future, experimental verification of two-stream instability in space will be an important topic. We also propose the following suggestions: (a) Three-dimensional PIC model can better capture the coupling effects between the magnetic and electric fields. PIC simulations using GPU parallel computing can more effectively elucidate wave–particle interactions in plasma environments, as GPUs enable the simultaneous advancement of numerous particles and the concurrent updating of electromagnetic fields. (b) The electron holes in phase space and the solitary wave pulse structures should be verified using Langmuir probes and transient electromagnetic sensors. The transient in-situ features should be extracted to validate the phase-space and real-space characteristics of the two-stream instability more clearly.

## ACKNOWLEDGEMENTS

**Fundings:** This research is supported by National Natural Science Foundation of China (92271113, 12411540222, 12481540165), the Fundamental Research Funds for Central Universities (2022CDJQY-003), the Chongqing Entrepreneurship and Innovation Support Programme for Overseas Returnees (CX2022004), and the Natural Science Foundation Project of Chongqing (CSTB2025NSCQ-GPX0725).
**Author contributions:** L.C. conceptualized this research. J.-K.S. performed the numerical simulation and wrote the manuscript. Y.L. and G.-J.W. provided constructive revisions on the structure and content of the manuscript. Z.-C.K. and S.-J.Z. collected the research materials and data. J.-J.M., D.-Z.L., and Y-X.Z. contributed to the revision of the manuscript. All authors approved the final manuscript.
**Competing interests:** The authors declare that they have no competing interests.

## DATA AVAILABILITY

The data that support the findings of this study are available upon reasonable request from the authors.